\documentclass{he_symp}
\usepackage{psfig}
\begin{document}

\title{Enlarging the Parameter Space and our Understanding \\
{\it or} The Radio Emission of Millisecond Pulsars}

\author{Michael Kramer\inst{1}}
\institute{University of Manchester, Jodrell Bank Observatory, UK}

\maketitle

\begin{abstract}
The understanding of pulsar radio emission demands a close look at all
regions of the huge parameter space present for pulsar observations.
In this review we concentrate on the space given by the range of
observed pulse periods, spanning four orders of magnitude. A
comparative study of the emission properties of millisecond pulsars
and normal pulsars, both at radio and high energy frequencies, promises
to shed light on the still poorly understood emission theory.
\end{abstract}

\section{Introduction}

The search for a theoretical framework that is able to explain the
radio emission of pulsars, has been largely unsuccessful. This is due
to the difficulty to include all the different phenomena and time
scales observed in radio pulsars into a working model. It is essential
to isolate the bigger picture, and one way of doing this, is to
enlarge the parameter space. Rather than making the picture even more
complicated by more observations that are difficult to explain, the
aim must be to study the emission properties by pushing the known
parameter space to new boundaries to provide solid
and obvious constraints.

The parameter space relevant for pulsar radio emission is
large. Obvious parameters are the period and its increase,
determining important values like magnetic field, or potential drop
above the surface. Further parameters are time resolution, frequency
coverage in the radio, but also from the radio regime to high energies,
eventually simultaneously, and sensitivity. A large neglected
part of the 
parameter space in pulsar studies is that of observing length
-- quite naturally as many time allocation committees do not appreciate
long-term projects. However, some
new phenomena will only be discovered, when sources -- not only
pulsars, that is --- are monitored for a  
long time span. In this review I will concentrate
on the period space, and the implication that can be derived from
studying emission of millisecond pulsars. For the other aspects, the
reader may be referred to other contributions (e.g.~Karastergiou et 
al.) or recent literature (e.g.~Kramer et al.~2002).

\section{Millisecond Pulsars}

Despite a manifold increase in the number of known pulsars,
the smallest period of any known pulsar is still that of PSR
B1937+21 discovered by Backer et al.~(1982) with 1.56 ms. 
As a consequence of their small periods millisecond pulsars
(MSPs)  are surrounded by magnetospheres which are 6 to 7 orders of
magnitude more compact than those of slower rotating pulsars. 
Inferred magnetic
fields close to the surface of MSPs are 3 to 4 orders weaker than in
normal pulsars. In contrast, charges at these regions experience an
accelerating potential similar to that of normal pulsars.  The impact
of the different environment on the emission processes in MSPs
magnetospheres has been a question addressed already shortly after
their discovery.  With the plethora of MSPs detected over the years, a
better
understanding of not only MSPs (as radio sources and tools) but slower
rotating (normal) pulsars as well. Recent addressing the radio
emission of pulsars are those by Kramer et
al.~(1998, Paper I) on spectra, pulse shapes and beaming fraction, by
Xilouris et al.~(1998, Paper II) on polarimetry of 24 MSPs, by Sallmen
(1998) and Stairs et al.~(1999) on multi-frequency polarimetry,
by Toscano et al.~(1998) on spectra of Southern MSPs, by Kramer et
al.~(1999b, Paper III) on multi-frequency evolution, and Kramer et
al.~(1999a, Paper IV) on profile instabilities of MSPs. 
A number of contributions can also be found in the proceedings
of IAU Colloquium 177.

\begin{figure}
\centerline{\psfig{file=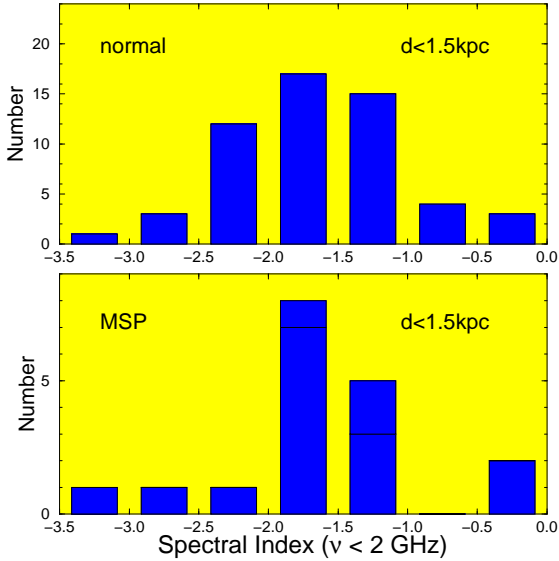,angle=-90,height=7.5cm} }

\caption{ \label{spectra}
Distribution of spectral indices for normal pulsars
and MSPs, measured in a frequency range below 2 GHz. 
Note that all data have been
derived from distance limited samples in order to avoid
selection effects (see text for details).}
\end{figure}

\subsection{Flux Density Spectra}

Prior to the investigations leading to Paper I it was commonly
believed that the spectra of millisecond pulsars were steeper than
those of normal pulsars.  It was demonstrated in Paper I that the
distribution of spectral indices for MSPs is in fact not significantly
different, finding an average index of $-1.76\pm0.14$ (Paper III). The
initial impression was due to a selection effect, since the first MSPs
were discovered in previously unidentified steep spectrum sources.
Results presented in
Paper III also suggest that most spectra can be represented by a
simple power law, i.e.~clear indications for a steepening at a few GHz
as known from normal pulsars are not seen. 

It is interesting to note that the current data suggest a similar mean
flux density spectrum but the distribution spectral indices seems to
be somewhat narrower for MSPs than for normal pulsars 
(Fig.~\ref{spectra}).

\begin{figure}
\centerline{\psfig{file=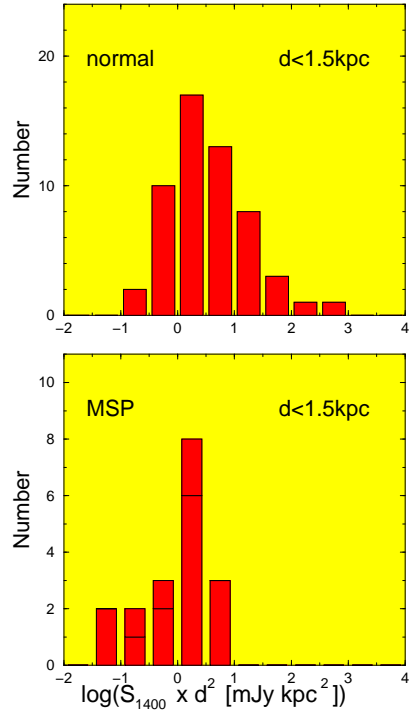,height=10cm} }

\caption{ \label{lum}
Luminosity of normal pulsars and MSPs derived from the flux
density measured at 1400 MHz. Note that all data have been
derived from distance limited samples in order to avoid
selection effects (see text for details).}
\end{figure}

\setcounter{footnote}{1}
\footnotetext{Upper row: MSPs (PSRs J0218+4218, J0621+1001, B1534+12,
J1640+2224, J1730$-$2304), lower row: normal pulsars (PSRs B1831$-$04,
B2045$-$16, B2110+27, B2016+28, B1826$-$17)}

\subsection{Radio Luminosity}

When comparing the spectra 
and radio luminosity of normal pulsars and MSPs, one
has to take extra care that selection effects
do not bias the result. The sample of
MSPs in the galactic plane is limited by scattering effects, resulting
in the discovery of mostly nearby MSPs. In contrast, normal pulsars
are also affected by scattering, but the longer period allows their
detection to larger distances. It is therefore useful to compare
only a distance limited sample that can be assumed to be sampled
adequately. A distance of 1.5 kpc is appropriate, as outlined in
Paper I. As a result one obtains that MSPs are slightly less luminous,
being also slightly less efficient in converting spin-down luminosity 
into radio emission (Fig.~\ref{lum})

Bailes et al.~(1997) pointed out that isolated MSPs are less luminous
than those in binary systems, pointing towards a possible relation
between radio luminosity and birth scenarios.  We have compared a
distance limited sample of normal and MSP and came to a similar result
with the MSPs as a whole being weaker sources than normal pulsars.

\begin{figure*}
\centerline{\psfig{file=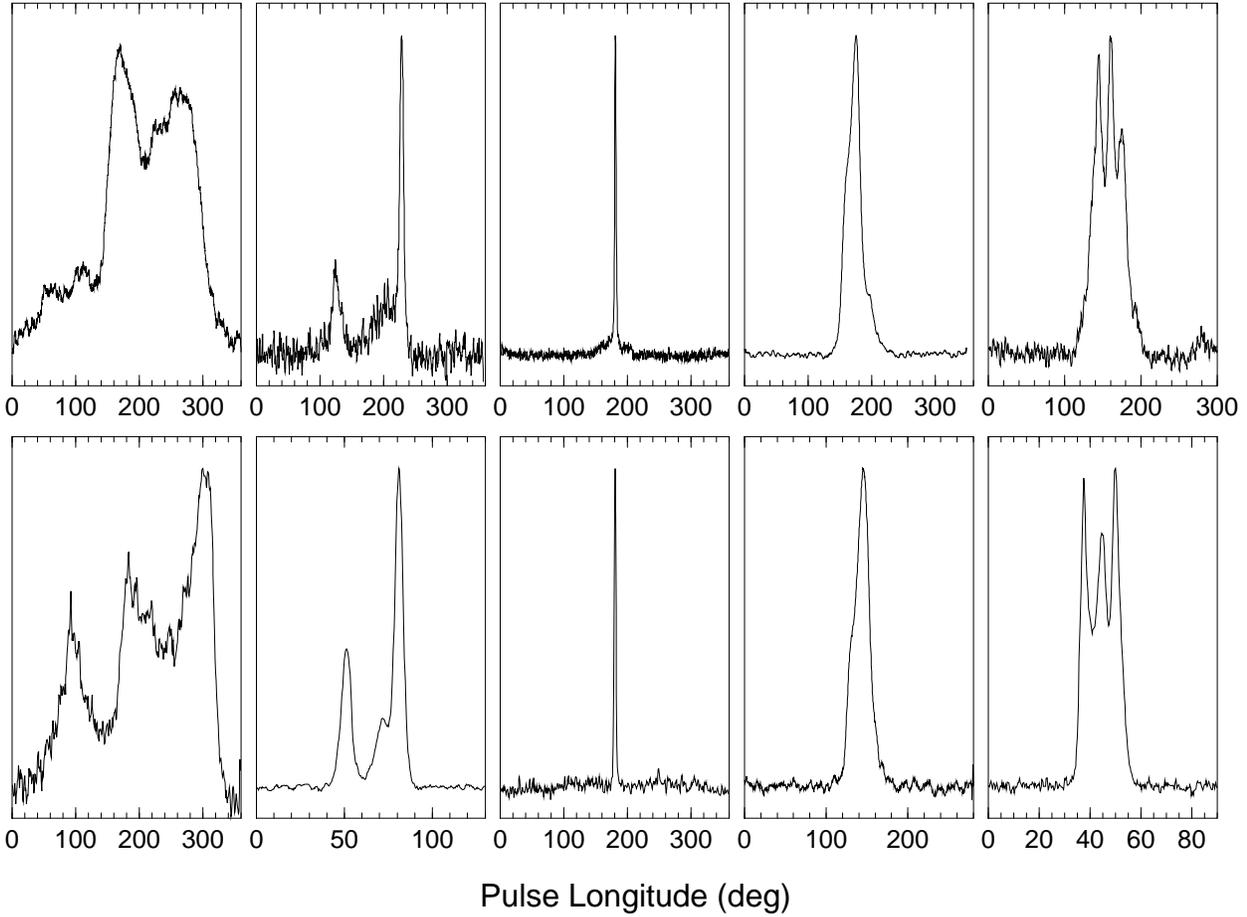,angle=-90,height=12cm} }

\caption{ 
\label{profiles}
Pulse profiles for a selected sample of MSPs and normal pulsars
(Paper I and EPN database). 
Note the similarity and make a guess which is which! See footnote$^1$
for the solution.}
\end{figure*}

\subsection{Pulse Profiles -- Complexity, Interpulses and Beaming Fraction}

For a long time it was also believed that MSP pulse profiles are more
complex than those of normal pulsars.  Using a large uniform sample of
profiles for fast and slowly rotating pulsars, we showed in Paper I
that the apparent larger complexity is due to the (typically) larger
duty cycle of MSPs. MSP profile can hence studied in greater detail,
facilitating the recognition of detailed structure. In fact, blowing
up normal pulsar profiles in a similar manner leads to very similar
profiles.  A quantitative proof is given in Paper I, while
Fig.~\ref{profiles} provides an illustration.

\begin{figure}
\centerline{\psfig{file=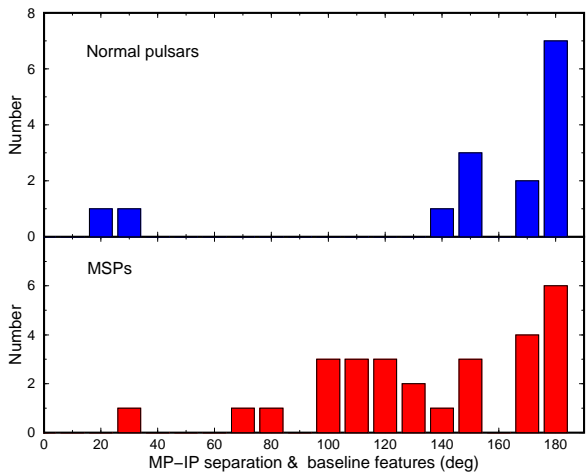,angle=-90,height=7cm} }

\caption{  Location of addition pulse feature across the pulse period for
  normal pulsars and MSPs. \label{features}}
\end{figure}

Despite this similarity, there is a profound difference to the
profiles of normal pulsars! Additional pulse features like
interpulses, pre- or post-cursor are much more common for MSPs. While
only $\sim2$\% of all normal pulsars are known to show such features,
we detect them for more than 30\% of all (field) MSPs. They also
appear at apparently random positions across the pulse period than we
see for normal pulsars (Fig.~\ref{features}). Their frequent occurrence and
location makes one wonder --- given the similarity of the main pulse
shapes otherwise --- whether these components are of the same origin
as the main pulse profile or whether other sources of emission
are responsible.

\begin{figure*}
\centerline{\psfig{file=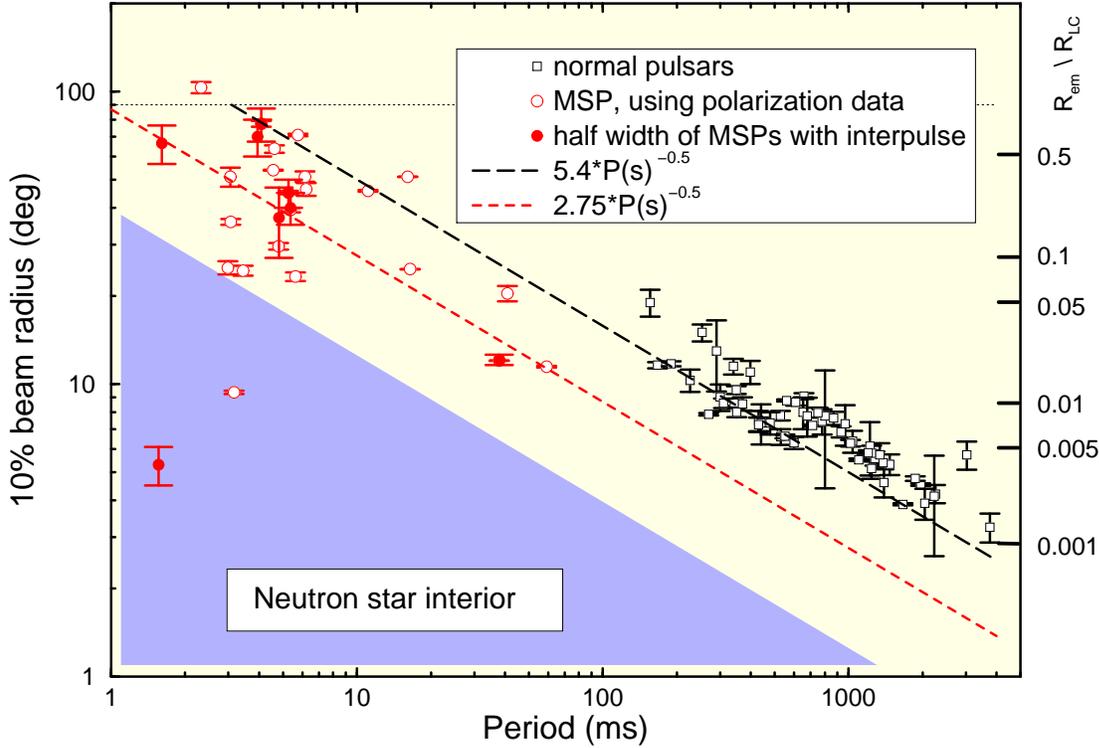,angle=-90,height=12cm} }

\caption{ \label{rho}
The beam radius, $\rho$, for normal pulsars
  and MSPs. MSPs do not follow the scaling law of normal pulsars (here Gould
  1994) but their beaming fraction is much smaller. For MSPs with interpulses
an ``inner'' relationship is indicated.}
\end{figure*}

One interpretation of these additional pulse features is related
to a  model first put forward for some young pulsars by
Manchester (1996), who interpreted some interpulses as the results of
cuts through a very wide cone. This is an interesting possibility also
for MSPs, since their beam width appears to be much smaller than
predicted from the scaling law derived for normal pulsars. The beam
width of normal pulsars, $\rho$, i.e.~the pulse width corrected for
geometrical effects (see Gil et al.~1984), follows a distinct $\rho
\propto P^{-0.5}$-law (e.g.~Rankin 1993, Kramer et al.~1994, Gould
1994). Using polarization information to determine the viewing
geometry and also applying statistical arguments, we calculated $\rho$
(at a 10\% intensity level) for MSPs and showed (Paper I) that they
are not only much smaller than the extrapolation of the known law to
small periods, but that -- under the assumption of dipolar magnetic
fields -- the emission of some MSPs seems to come even from within the
neutron star --- a really disturbing result! While we discuss the
possibility of non-dipolar fields and the used the polarization
information below, one explanation would be that (perhaps below a
critical period) the emission beam does not fill the whole open field
line region (``unfilled beam''). The situation improves a bit when we
consider the additional pulse features as regular parts of the pulse
profile (Fig.~\ref{rho}).  In fact, those MSPs with interpulses may indicate
an additional inner scaling parallel to that known for normal pulsars,
which could be a result of unfilled beams.

\begin{figure}

\centerline{\psfig{file=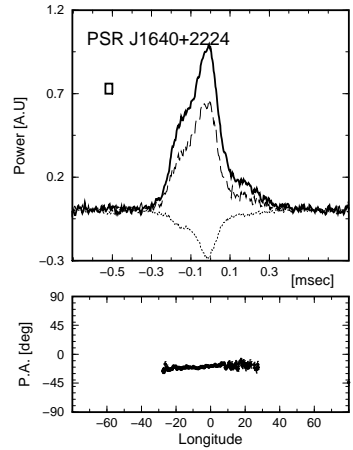,height=7cm} }

\caption{ Profile of  PSR J1640+2224 as an example for a 
MSP exhibiting a flat PA swing.\label{flat}}
\end{figure}

Another explanation of the additional pulse features could be the
possibility that so-called ``outer gap'' emission is responsible (see
Paper II).  A number of MSPs are detected as X-ray sources (see
contribution by Becker) and those with magnetospheric emission are
probably detected by emission created in the outer magnetosphere, 
i.e.~~``outer gaps'' (see contribution by Romani). As MSPs have very
compact magnetospheres, the usual location of radio emission in some
distance to the star could well coincide with the location of outer
gaps, enabling the detection of ``classical'' polar-cap radio emission
and that of outer gaps at the same time. This is similar to the
circumstances probably responsible for the additional high-frequency
components seen in the Crab pulsar (Moffett \& Hankins 1996).

\begin{figure}
\centerline{ \psfig{file=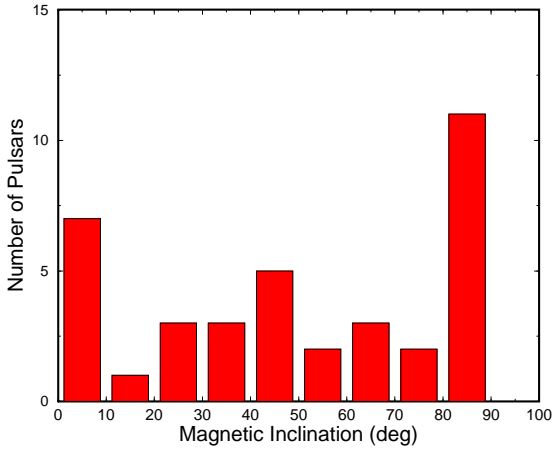,angle=-90,height=7cm} }

\caption{ Ddistribution of magnetic inclination angles derived from RVM fits.
\label{incl}}
\end{figure}

Finally, it should be pointed out 
that the much smaller beam width has consequences for
population studies, which usually utilise the $\rho-P$ scaling as
found for normal pulsars. The failure of this law leads to an
overestimated beaming fraction and an underestimation of the birth
rate of recycled pulsars (see Paper I).

\begin{figure}
\centerline{\psfig{file=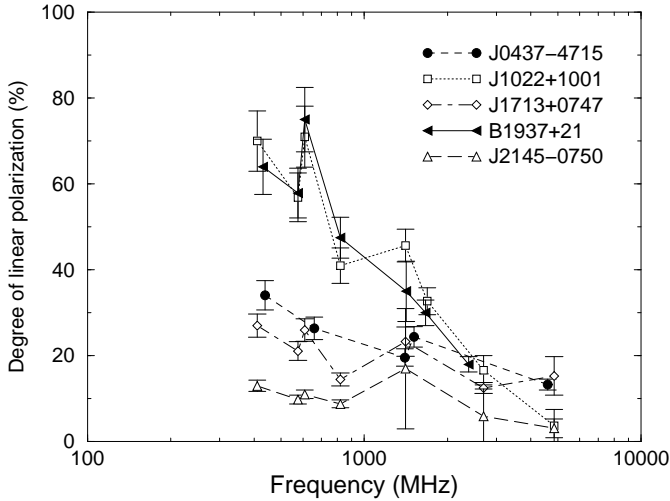,height=7cm} }

\caption{Degree of polarization for some MSPs as a function of frequency.
\label{depol}}
\end{figure}

\subsection{Polarization Properties}

The radio emission of MSPs shows all polarization features known from
normal pulsars, i.e.~circular polarization which is usually associated
with core components, linear polarization which is usually associated
with cone components, and also orthogonal polarization modes (see
Paper II, Sallmen 1998, Stairs et al.~1999). Despite the qualitative
similarities, the position angle (PA) swing is often strikingly
different.  While normal pulsars show typically a {\sf S}-like swing,
which is interpreted within the rotating vector model (RVM;
Radhakrishnan \& Cooke 1969), the PAs of many MSPs often appear flat
(see e.g.~Fig.~\ref{flat}). This could be interpreted in terms of non-dipolar
fields, but Sallmen (1998) noted that larger beam radii lead to a
larger probability for outer cuts of the emission cones, i.e.~ flatter
PA swings according to the RVM.  Although one should bear in mind the
limitations of the $\rho$-scaling law and another {\em caveat}
discussed later, this interpretation justifies the geometrical
interpretation of the data.  Magnetic inclination angles derived
from RVM fits are important for binary evolution models and
determinations of the companion mass (Fig.~ref{incl}).

\begin{figure}
\centerline{\psfig{file=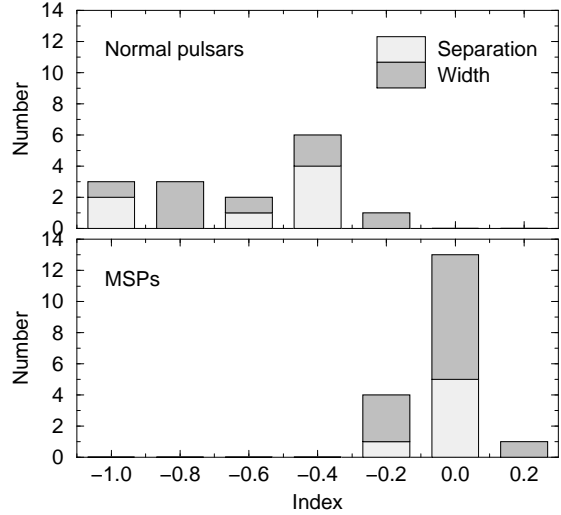,height=7cm} }

\caption{ Power law index of profile narrowing with frequency 
(see Paper III for details). \label{narrow}}
\end{figure}

\subsection{Frequency Evolution}

The radio properties of normal pulsars show a distinct frequency
evolution, i.e.~with increasing frequency the profile narrows, outer
components tend to dominate over inner ones, and the emission
depolarises. The emission of MSPs, which at intermediate frequencies
tends to be more polarized than that of normal pulsars (Paper II),
also depolarises at high frequencies (Fig.~\ref{depol}; Paper III).
Simultaneously, the profile width hardly changes or remains constant
(see Fig.~\ref{narrow}).  This
puts under test attempts to link both effects to the same physical
origin (i.e.~birefringes). In fact, many profiles also exhibit the
same shape at all frequencies, while others evolve in an unusual way,
i.e.~the spectral index of inner components is not necessarily
steeper, so that a systematic behaviour as seen for normal pulsars is
hardly observed. This can be understood in terms of a compact emission
region, an assumption further supported by a simultaneous arrival of
the profiles at all frequencies. We emphasize that we have not
detected any evidence for the existence of non-dipolar fields (Paper
III).

\subsection{Profile and Polarization Instabilities}

The amazing stability with time of MSP profiles has enabled high
precision timing over the years. However, in Paper IV we discussed the
surprising discovery that a few MSPs do show profile changes caused by
an unknown origin.  These time scales of these profile instabilities 
are inconsistent with the known mode-changing. In particular, PSR
J1022+1001 exhibits a narrow-band profile variation never seen before
(Paper IV), which could, however, be the result of magnetospheric
scintillation effects described by Lyutikov (2001). With
the pulse shape the polarization usually changes as well, and hence
this effect is possibly related to phenomena which we discovered in
Paper II. Some pulsars like PSR J2145--0750 (Paper II) or PSR
J1713--0747 (Sallmen 1998) show occasionally a profile which is much
more polarized than their usual pulse shape. In the case of PSR
J2145--0750, the PA changes from some distinct (though not {\sf
S}-like) swing to some very flat curve. This is a strong indication
that some of the flat PA swings discussed above may not be of simple
geometrical origin alone.

\section{Summary and Conclusions}

The emission properties of millisecond pulsars are in many respects
similar to those of slowly rotating pulsars. However, there are
a few remarkable differences like additional profile components
and a very week frequency evolution for most MSPs, which can be
attributed to the smaller, compact magnetosphere. The additional
pulse components may be representatives of outer gap emission,
providing an interesting link to the X-ray properties of MSPs.
This motivates a close inspection of their joint radio and high
energy characteristics. 
Results of this work in progress will be presented
elsewhere.

\begin{acknowledgements}
The author gratefully acknowledges the support by the Heraeus foundation
and thanks the organizers for a wonderful meeting.
\end{acknowledgements}
   
% Example list of References

\end{document}